\documentclass[aps,prb,10pt,longbibliography,superscriptaddress,twocolumn]{revtex4-2}
\usepackage[pdftex]{graphicx}
\usepackage{mathtools,amsmath,amssymb}
\usepackage{multirow}
\usepackage{xcolor}
\usepackage{hyperref}
\usepackage{bm}
\usepackage{physics}

\renewcommand{\Re}{\operatorname{Re}}
\renewcommand{\Im}{\operatorname{Im}}

\newcommand{\rmd}{\mathrm{d}}

\begin{document}
\title{Emergent-gravity Hall effect from quantum geometry}

\author{Hiroki Yoshida}
\affiliation{Department of Physics, Institute of Science Tokyo, 2-12-1 Ookayama, Meguro-ku, Tokyo 152-8551, Japan}
\author{Takehito Yokoyama}
\affiliation{Department of Physics, Institute of Science Tokyo, 2-12-1 Ookayama, Meguro-ku, Tokyo 152-8551, Japan}

\date{\today}

\begin{abstract}
    We theoretically propose a Hall effect driven by effective gravitational fields arising from quantum geometry. We identify four mechanisms for this ``emergent-gravity Hall effect'': real-space gravity, momentum-space gravity, gravitational anomalous velocity, and gravitational Lorentz force, all of which are described by Christoffel symbols in real, momentum, or time spaces. Based on the semiclassical theory, we construct a unified theoretical framework to systematically investigate how emergent gravity in these spaces affects transport phenomena. We demonstrate these effects through model calculations and clarify the conditions under which a finite Hall response can arise. Our findings open a new avenue for exploring gravitational effects in quantum systems.
\end{abstract}

\maketitle

\textit{Introduction}. Transport phenomena in solids are profoundly influenced by the geometry and topology of electronic wavefunctions. A prime example is the anomalous Hall effect~\cite{karplus_hall_1954,smit_spontaneous_1955,smit_spontaneous_1958,berger_side-jump_1970,berry_quantal_1984,ye_berry_1999,fang_anomalous_2003}, where it was found that the Berry curvature in momentum space --a geometrical quantity associated with the underlying Bloch states-- plays a crucial role in the transverse motion of electrons under an electric field. In addition to the Berry curvature, the quantum metric --another fundamental geometrical quantity of Bloch states-- has also been of central importance~\cite{provost_riemannian_1980,tan_experimental_2019,gianfrate_measurement_2020,klees_microwave_2020,yu_experimental_2020,gao_field_2014,gao_quantum_2023,Nagaosa_review_2025}. In particular, its momentum derivative constitutes the Christoffel symbol, representing momentum-space gravity~\cite{gao_field_2014,smith_momentum-space_2022,mehraeen_quantum_2025,fontana_quantum_2025}.

Attention has also shifted toward real-space geometry, particularly in systems with nontrivial spin textures. One notable development is the discovery of the topological Hall effect~\cite{tatara_chirality-driven_2002,bruno_topological_2004} induced by scalar spin chirality, which leads to an emergent electromagnetic field acting on the electrons~\cite{Nagaosa_Tokura_2012}. These advances underscore the growing interest in understanding how spatial geometry and spin structure can give rise to novel transport responses. This raises a natural question: are there emergent gravitational fields acting on electrons? 

In this work, we uncover a new type of Hall effect due to effective gravitational fields arising from quantum geometry, which can be described by the Christoffel symbols. In analogy with the term ``emergent electromagnetic field", we refer to this as an emergent gravitational field, and the associated Hall effect as the ``emergent-gravity Hall effect".

We theoretically predict this phenomenon by extending the semiclassical wavepacket dynamics to include interband transitions beyond the adiabatic approximation. This approach allows the Christoffel symbols to emerge naturally in the equation of motion of the wavepacket and enables a unified analysis of transport phenomena governed by quantum geometry in real, momentum, time, and parameter spaces. 
With this approach, we develop four mechanisms for this emergent-gravity Hall effect: real-space gravity, momentum-space gravity, gravitational anomalous velocity, and gravitational Lorentz force.
We demonstrate these mechanisms by model calculations and clarify the conditions for a finite Hall response.

\textit{General theory}. We begin by deriving the equation of motion, incorporating nonadiabatic time evolution. In Ref.~\cite{sundaram_wave-packet_1999}, a general Lagrangian for an electron with spatial, momentum, and time dependencies is provided. In addition to these, we introduce an arbitrary parameter $\lambda$, which depends solely on time. This parameter can, for example, represent a strain in the system. Following the same procedure of derivation, we obtain the Lagrangian of the system as~\cite{sundaram_wave-packet_1999},
\begin{widetext}
    \begin{align}
        L(\bm{r}_c,\dot{\bm{r}}_c,\bm{q}_c,\dot{\bm{q}}_c,t)&=-\varepsilon-\hbar\dot{\bm{q}}_c\cdot\bm{r}_c+\hbar\dot{\bm{r}}_c\cdot\left\langle u\middle|i\frac{\partial u}{\partial \bm{r}_c}\right\rangle+\hbar\dot{\bm{q}}_c\cdot\left\langle u\middle|i\frac{\partial u}{\partial \bm{q}_c}\right\rangle+\hbar\dot{\lambda}\left\langle u\middle|i\frac{\partial u}{\partial \lambda}\right\rangle+\hbar\left\langle u\middle|i\frac{\partial u}{\partial t}\right\rangle,\label{eq:General_Lagrangian}
    \end{align}
\end{widetext}
where $\varepsilon$ is the energy of a wavepacket, $\bm{r}_c$ and $\bm{q}_c$ are centers of the wavepacket in real and reciprocal spaces, respectively, and $\left|u\right\rangle\coloneqq \left| u(\bm{r}_c,\bm{q}_c,t)\right\rangle$ is the periodic part of the Bloch state. In the original work~\cite{sundaram_wave-packet_1999}, only a single band
is considered. Here, we consider the time evolution of the quantum state beyond the adiabatic approximation.

We assume that the system is initially in the $n$-th eigenstate of the Hamiltonian. For notational simplicity, we omit the index $c$. Under the assumption that the eigenstates vary slowly in time compared to the band gap, the state $|u\rangle$ can be expanded as~\cite{Nagaosa_review_2025,xiao_berry_2010} (see Supplemental Material~\cite{supplementary} for details)
\begin{align}
    \left\lvert u\right\rangle=&\left\lvert u_n(\bm{r},\bm{q},\lambda,t)\right\rangle\nonumber\\
    &+\hbar\sum_{m\neq n} \frac{\dot{\bm{r}}\cdot\bm{A}^r_{mn}+\dot{\bm{q}}\cdot\bm{A}^q_{mn}+\dot{\lambda}A^{\lambda}_{mn}+A^t_{mn}}{\varepsilon_{mn}}\nonumber\\
    &\hspace{4cm}\times\left\lvert u_m(\bm{r},\bm{q},\lambda,t)\right\rangle,\label{eq:u-expand}
\end{align}
where the Berry connections in each space are defined as $\bm{A}^X_{mn,i}\coloneqq \left\langle u_m\middle|i\frac{\partial u_n}{\partial X_i}\right\rangle$ with $X=r,q$, $A^Y_{mn}\coloneqq \left\langle u_m\middle\lvert i\frac{\partial u_n}{\partial Y}\right\rangle$ with $Y=\lambda,t$, while $\varepsilon_{mn}=\varepsilon_m-\varepsilon_n$. This expression represents the first-order correction to the adiabatic approximation. By substituting this expansion into Eq.~\eqref{eq:General_Lagrangian}, we obtain a Lagrangian that incorporates nonadiabatic processes. We assume that the time variation is sufficiently slow and neglect all terms involving third- and higher-order time derivatives. The resulting Lagrangian can then be written as
\begin{align}
    L_n&=-\varepsilon_n-\hbar\dot{\bm{q}}\cdot\bm{r}+\hbar A^q_{nn,j}\dot{q}_j+\hbar A^r_{nn,j}\dot{r}_j+\hbar A^{\lambda}_{nn}\dot{\lambda}+\hbar A^t_{nn}\nonumber\\
    &+\hbar^2\frac{\dot{q}_j}{2}\left(G_{ji}^{qr}\dot{r}_i+G^{qq}_{ji}\dot{q}_i+G^{q\lambda}_{j}\dot{\lambda}+G^{qt}_{j}\right)\nonumber\\
    &+\hbar^2\frac{\dot{r}_j}{2}\left(G_{ji}^{rr}\dot{r}_i+G^{rq}_{ji}\dot{q}_i+G^{r\lambda}_{j}\dot{\lambda}+G^{rt}_{j}\right)\nonumber\\
    &+\hbar^2\frac{\dot{\lambda}}{2}\left(G_{\phantom{j}i}^{\lambda r}\dot{r}_i+G^{\lambda q}_{\phantom{j}i}\dot{q}_i+G^{\lambda\lambda}\dot{\lambda}+G^{\lambda t}\right)\nonumber\\
    &+\hbar^2\frac{1}{2}\left(G^{tr}_{\phantom{j}i}\dot{r}_i+G^{tq}_{\phantom{j}i}\dot{q}_i+G^{t\lambda}\dot{\lambda}+G^{tt}\right),\label{eq_Lagrangian_n}
\end{align}
where $\varepsilon_n$ is the energy of the $n$th band, $G^{XY}_{ij}\coloneqq 2\sum_{m\neq n}\Re\left[\frac{A^X_{nm,i}A^Y_{mn,j}}{\varepsilon_{mn}}\right],\ (X,Y=r,q)$ and others are defined in a similar manner. These quantities are the so-called weighted quantum metric~\cite{jain_anomalous_2025}. From this Lagrangian, the equations of motion for an electron can be derived using the Euler-Lagrange equation. A straightforward calculation leads to
\begin{align}
    \dot{q}_a =&-\frac{1}{\hbar}\frac{\partial \varepsilon_n}{\partial r_a}+\Omega^{rr}_{ai}\dot{r}_i+\Omega^{rq}_{ai}\dot{q}_i+\Omega^{r\lambda}_{a}\dot{\lambda}+\Omega^{rt}_{a}\nonumber\\
    &-\hbar\Gamma^{rrr}_{a,ij}\dot{r}_i\dot{r}_j-2\hbar\Gamma^{rrq}_{a,ij}\dot{r}_i\dot{q}_j-2\hbar\Gamma^{rr\lambda}_{a,i\phantom{j}}\dot{r}_i\dot{\lambda}-2\hbar\Gamma^{rrt}_{a,i\phantom{j}}\dot{r}_i\nonumber\\
    &-\hbar\Gamma^{rqq}_{a,ij}\dot{q}_i\dot{q}_j-2\hbar\Gamma^{rq\lambda}_{a,ij}\dot{q}_i\dot{\lambda}-2\hbar\Gamma^{rqt}_{a,i\phantom{j}}\dot{q}_i\nonumber\\
    &-\hbar\Gamma^{r\lambda\lambda}_{a,}\dot{\lambda}\dot{\lambda}-2\hbar\Gamma^{r\lambda t}_{a,}\dot{\lambda}\nonumber\\
    &-\hbar\Gamma^{rtt}_{a,}-\hbar\left(G^{rr}_{ai}\ddot{r}_i+G^{rq}_{ai}\ddot{q}_i+G^{r\lambda}_{a}\ddot{\lambda}\right),\label{eq:General_EOM_q}\\
    \dot{r}_a =&\frac{1}{\hbar}\frac{\partial \varepsilon_n}{\partial q_a}-\Omega^{qr}_{ai}\dot{r}_i-\Omega^{qq}_{ai}\dot{q}_i-\Omega^{q\lambda}_{a}\dot{\lambda}-\Omega^{qt}_{a}\nonumber\\
    &+\hbar\Gamma^{qrr}_{a,ij}\dot{r}_i\dot{r}_j+2\hbar\Gamma^{qrq}_{a,ij}\dot{r}_i\dot{q}_j+2\hbar\Gamma^{qr\lambda}_{a,i\phantom{j}}\dot{r}_i\dot{\lambda}+2\hbar\Gamma^{qrt}_{a,i\phantom{j}}\dot{r}_i\nonumber\\
    &+\hbar\Gamma^{qqq}_{a,ij}\dot{q}_i\dot{q}_j+2\hbar\Gamma^{qq\lambda}_{a,i\phantom{j}}\dot{q}_i\dot{\lambda}+2\hbar\Gamma^{qqt}_{a,i\phantom{j}}\dot{q}_i\nonumber\\
    &+\hbar\Gamma^{q\lambda\lambda}_{a,}\dot{\lambda}\dot{\lambda}+2\hbar\Gamma^{q\lambda t}_{a,}\dot{\lambda}\nonumber\\
    &+\hbar\Gamma^{qtt}_{a,}+\hbar\left(G^{qr}_{ai}\ddot{r}_i+G^{qq}_{ai}\ddot{q}_i+G^{q\lambda}_{a}\ddot{\lambda}\right)\label{eq:General_EOM_r},
\end{align}
where $\Omega^{XY}_{ij}\coloneqq\frac{\partial A^Y_{nn,j}}{\partial X_i}-\frac{\partial A^X_{nn,i}}{\partial Y_j}=-\Omega^{YX}_{ji}$ is the Berry curvature.
An important point here is that the inclusion of nonadiabatic processes yields correction terms in the form of Christoffel symbols, defined from the weighted quantum metric as
\begin{align}
    \Gamma^{abc}_{i,jk}&\coloneqq \frac{1}{2}\left(\frac{\partial G^{ab}_{ij}}{\partial c_k}+\frac{\partial G^{ac}_{ik}}{\partial b_j}-\frac{\partial G^{bc}_{jk}}{\partial a_i}\right).
\end{align}
By definition, $\Gamma^{abc}_{i,jk}=\Gamma^{acb}_{i,kj}$ holds. 
In analogy with the concept of the emergent electromagnetic field~\cite{volovik_linear_1987,Nagaosa_Tokura_2012,nagaosa_topological_2013}, we refer to these quantities as emergent gravitational fields. Below, we show how this emergent gravity manifests in electronic transport properties.

Using the Boltzmann equation with the relaxation time approximation, we then derive the electron distribution function to arbitrary order in the relaxation time. Combining the equation of motion derived above with the distribution function allows us to compute the electric current in the system. Note that, as pointed out in Ref.~\cite{xiao_berry_2005}, when the Berry curvature in mixed spaces, such as $\Omega^{rq}$, is nonzero, the density of states is modified. In the following, we consider only the case where $\Omega^{rq}=0$.

Below, we consider three cases and derive the emergent-gravity Hall effect caused by emergent gravity in 1.~real space, 2.~momentum space, and 3.~time-domain. 

\textit{1.~Real space}. Here we consider the effects of geometrical quantities in real space. We assume a two-band Hamiltonian of the form 
\begin{align}
    H(\bm{r},\bm{q})&=h_0(\bm{q})\sigma_0+\bm{h}(\bm{r})\cdot\boldsymbol{\sigma}\label{eq:R_2band_Hamiltonian},
\end{align}
where $\sigma_0$ is an identity matrix and $\boldsymbol{\sigma}$ is a vector of Pauli matrices. In this case, Bloch states depend only on $\bm{r}$, implying that the terms in Eqs.~\eqref{eq:General_EOM_q} and \eqref{eq:General_EOM_r} involving $q$, $t$ and $\lambda$ vanish, and the equations of motion reduce to
\begin{align}
    \dot{q}_a&=-\frac{1}{\hbar}\frac{\partial V}{\partial r_a}+\frac{1}{\hbar}\Omega^{rr}_{ai}\frac{\partial h_0}{\partial q_i}-\frac{1}{\hbar}\Gamma^{rrr}_{a,ij}\frac{\partial h_0}{\partial q_i}\frac{\partial h_0}{\partial q_j},\label{eq:Xonly_EOM_k}\\
    \dot{r}_a&=\frac{1}{\hbar}\frac{\partial h_0}{\partial q_a}\label{eq:Xonly_EOM_r},
\end{align}
where an external potential $V(\bm{r})$ is introduced. In the following, we consider an electric field applied to the system, defined as $-e E_a = -\frac{\partial V}{\partial r_a}$. For this system, the distribution function can be obtained from the Boltzmann equation with the relaxation time approximation:
\begin{align}
    f(\varepsilon(\bm{r},\bm{q}))&=f_0-\tau\left(\dot{\bm{q}}\cdot\frac{\partial f}{\partial \bm{q}}+\dot{\bm{r}}\cdot\frac{\partial f}{\partial\bm{r}}\right),
\end{align}
where $f_0$ is the Fermi distribution function without external perturbation, $f$ is the modified distribution function, and $\tau$ is a relaxation time. We note that even though we consider nonadiabatic processes here, we only incorporate the lowest order of nonadiabaticity and the time scale of the system is much longer than the relaxation process. Therefore, the relaxation time approximation is expected to be a good approximation. Beyond the relaxation-time approximation, a more complete treatment of the collision integral may introduce additional contributions such as side-jump and skew-scattering mechanisms~\cite{Sinitsyn_review}, whose analysis is left for future work. The electric current is given by $j_a=-e\int Df(\varepsilon)\dot{r}a$, where the integral denotes $\int \frac{\mathrm{d}^dq}{(2\pi)^d}$, with $d$ being the spatial dimension of the system, and $D$ is the density-of-states factor. Here, we focus on terms that are first order in the external electric field. The contribution to the current that is first order in the electric field and second order in the relaxation time is given by
\begin{align}
    &j^{(\tau^2)}_a=\frac{e^2\tau^2}{\hbar^3}\nonumber\\
    &\ \times\int\left[\left\{\Omega^{rr}_{cj}-\Gamma^{rrr}_{c,jk}\frac{\partial h_0}{\partial q_k}\right\}\frac{\partial h_0}{\partial q_j}\frac{\partial h_0}{\partial q_a}\frac{\partial^2 h_0}{\partial q_b\partial q_c}\frac{\partial f_0}{\partial \varepsilon_n}\right]E_b.\label{eq:Ronly_j_E1}
\end{align}
Terms proportional to $\tau^1$ also appear, but they are not considered here because they do not contribute to Hall responses (see Supplemental Material~\cite{supplementary} for details of the calculation). This current contains both symmetric and antisymmetric components under interchange of the indices $a$ and $b$. We define the conductivity tensor by $j^{(\tau^2)}_a=\sigma_{ab}E_b$. In addition to the term involving $\Omega^{rr}$, which gives rise to the topological Hall effect, we obtain a contribution proportional to $\Gamma$. Only the antisymmetric part of the conductivity tensor, $\sigma^{\mathrm{asym}}_{ab}\coloneqq (\sigma_{ab}-\sigma_{ba})/2$, contributes to the Hall effect; it is given by
\begin{align}
    \sigma^{\mathrm{asym}}_{ab}&\coloneqq \frac{e^2\tau^2}{2\hbar^3} \Gamma^{rrr}_{c,jk}\nonumber\\
    &\quad\times\int\left(\frac{\partial h_0}{\partial q_b}\frac{\partial^2h_0}{\partial q_a\partial q_c}-\frac{\partial h_0}{\partial q_a}\frac{\partial^2h_0}{\partial q_b\partial q_c}\right)\frac{\partial h_0}{\partial q_k}\frac{\partial h_0}{\partial q_j}\frac{\partial f_0}{\partial \varepsilon_n}.\label{eq:sigma_E1_asym}
\end{align}
This term is proportional to the Christoffel symbols and describes the Hall response induced by the emergent gravitational field. This effect requires that the energy is not an even function of $\bm{q}$. To distinguish it from the topological Hall effect, one may consider spin configurations with vanishing scalar spin chirality.

Before going into model calculations, we give expressions of the weighted quantum metric and Christoffel symbols for general two-band systems in Eq.~\eqref{eq:R_2band_Hamiltonian}. With $h\coloneqq \left|\bm{h}(\bm{r})\right|$, the weighted quantum metric and the Christoffel symbol in real space are given by
\begin{align}
    G^{rr}_{ij}&=\frac{1}{4h^3}\left(\partial_ih_{\mu}\partial_jh_{\mu}-\partial_ih\partial_jh\right),
\end{align}
and 
\begin{align}
    \Gamma^{rrr}_{a,ij}&=-\frac{3}{8h^4}\left(\partial_{a}h_{\mu}\partial_ih_{\mu}\partial_jh+\partial_{a}h_{\mu}\partial_ih\partial_jh_{\mu}-\partial_{a}h\partial_ih_{\mu}\partial_jh_{\mu}\right)\nonumber\\
    &+\frac{1}{4h^3}\left(\partial_ah_{\mu}\partial_i\partial_jh_{\mu}-\partial_ah\partial_i\partial_jh\right)+\frac{3}{8h^4}\partial_ah\partial_ih\partial_jh,\label{eq:2level_Gamma}
\end{align}
respectively. In the expressions above, summation over $\mu = x, y, z$ is taken. Note that if $h_{\mu}(\bm{r})=\pm h_{\mu}(-\bm{r})$, $\Gamma^{rrr}$ is an odd function of position. Below, we calculate $\Gamma^{rrr}$ for a. one-dimensional domain-wall and b. two-dimensional skyrmion structures as examples.

\textit{a.~Magnetic domain-wall}. Although the Hall effect is absent in one-dimensional systems, we examine the behavior of the emergent gravitational field $\Gamma^{rrr}$ by analyzing a one-dimensional domain-wall structure as an illustrative example.

The spin texture of the domain wall is described by the exchange field~\cite{shibata_brief_2011}
\begin{align}
    \bm{h}(x) &=\left(0,h\sin\theta,h\cos\theta\right),
\end{align}
with
\begin{align}
    \theta &= 2\tan^{-1}\exp \left(-\frac{x}{a}\right).
\end{align}
Here, $h$ is a constant. Therefore, from the general expression in Eq.~\eqref{eq:2level_Gamma}, the Christoffel symbol for a one-dimensional domain wall reduces to
\begin{align}
    \Gamma^{rrr}_{x,xx}&=\frac{1}{4h^3}\frac{\partial^2h_{\mu}}{\partial x^2}\frac{\partial h_{\mu}}{\partial x}=\frac{1}{4h}\frac{\partial \theta}{\partial x}\frac{\partial^2\theta}{\partial x^2}.
\end{align}

\begin{figure}[t]
    \begin{center}
        \includegraphics[width=\columnwidth]{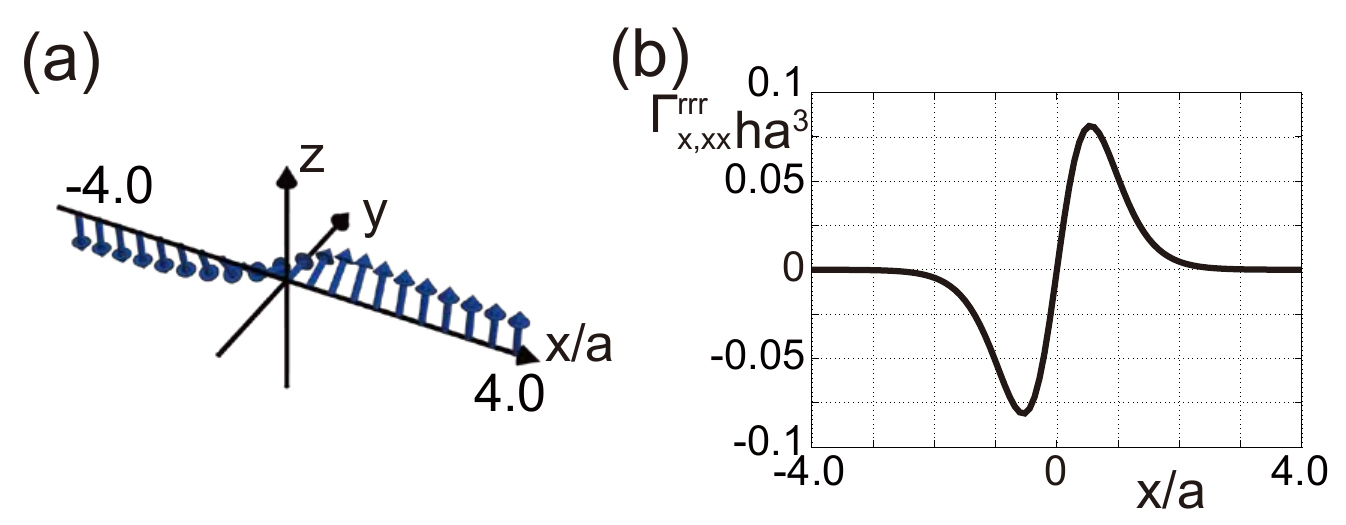}
        \caption{1D domain-wall structure and the Christoffel symbol $\Gamma^{rrr}_{x,xx}$ as a function of position $x$.
        (a) Domain-wall spin configuration. Blue arrows represent the spatially varying spin texture.
        (b) The Christoffel symbol $\Gamma^{rrr}_{x,xx}(x)$. In this system, $\Gamma^{rrr}_{x,xx}(x)$ is an odd function of $x$: $\Gamma^{rrr}_{x,xx}(x) = -\Gamma^{rrr}_{x,xx}(-x)$.}
        \label{fig:1DDW}
    \end{center}
\end{figure}

In Fig.~\ref{fig:1DDW}, we plot $\Gamma^{rrr}_{x,xx}(x)$ together with the spin configuration as a function of position $x$. This figure shows that the emergent gravitational field is locally nonzero, even in a simple spin configuration that does not possess scalar spin chirality. The emergent gravitational field becomes large where the spin texture varies rapidly in space. However, since $\Gamma^{rrr}_{x,xx}(x)$ is an odd function of $x$, its spatial average vanishes.

\textit{b.~Skyrmion}. We now turn to the main topic of this paper: demonstration of emergent-gravity Hall effect. As a representative example, we consider a two-dimensional single skyrmion described by~\cite{Belavin_Metastable_1975}
\begin{align}
    \bm{h}(x,y)&=\frac{h}{1+|u|^2}\left(2\Re\left(u\right),2\Im\left(u\right),1-|u|^2\right),\label{eq:sky_h}\\
    u(x,y)&\coloneqq\frac{ia}{x-iy}\label{eq_sky_u}.
\end{align}
For this system, we calculate the emergent-gravity Hall conductivity given by Eq.~\eqref{eq:sigma_E1_asym}. Since the integrand of Eq.~\eqref{eq:sigma_E1_asym} contains five derivatives with respect to momentum, the antisymmetric conductivity $\sigma^{\mathrm{asym}}_{ab}$ vanishes if the system satisfies the symmetry $h_0(\bm{q}) = h_0(-\bm{q})$. To obtain a nonzero conductivity, we need $h_0$ which is not an even function of $\bm{q}$. Thus, we consider a noncentrosymmetric magnet and assume $h_0=\hbar^2\left|\bm{q}\right|^2/(2m)+\alpha q_x^3$, where $m$ is the electron mass. The calculated conductivity is presented together with the skyrmion spin texture in Fig.~\ref{fig:2DSK} (for details of the calculation, see Supplemental Material~\cite{supplementary}). We find a finite emergent-gravity Hall conductivity with a maximum magnitude of $10^{-4}e^2/\hbar$. This value is three orders of magnitude smaller than the conductivity per skyrmion estimated from the measured topological Hall resistance of the skyrmion lattice reported in Ref.~\cite{yu_observation_2024}. In experiments, the spatially averaged current is typically measured. However, for a single isotropic skyrmion, the spatial average of the conductivity tensor vanishes, since all six independent components of $\Gamma^{rrr}$ are odd functions of $\bm{r}$. This issue can be circumvented by applying an inhomogeneous electric field so that the spatial average of the current becomes finite or using an inversion-symmetry-broken skyrmion lattice with $\bm{h}(\bm{r})\neq\pm \bm{h}(-\bm{r})$.

\begin{figure}[t]
    \begin{center}
        \includegraphics[width=\columnwidth]{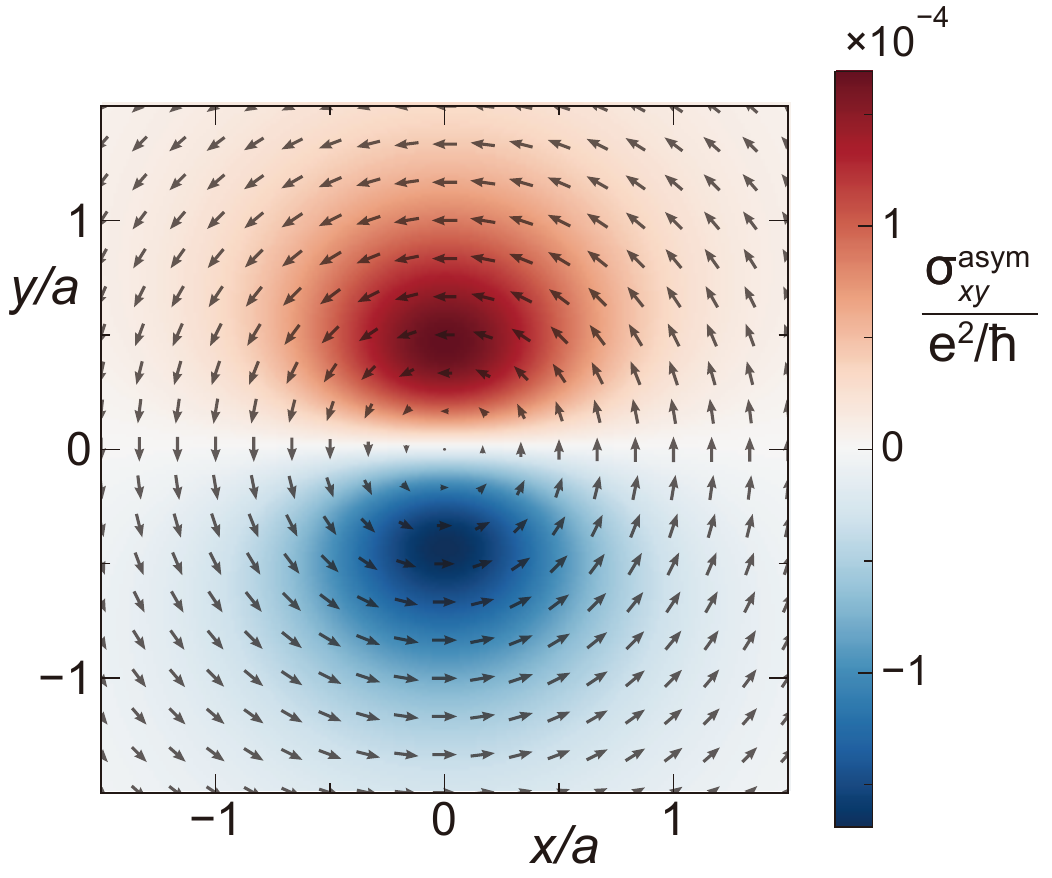}
        \caption{Emergent-gravity Hall conductivity at zero temperature in real space. The arrows represent the skyrmion spin texture. Parameters are $q_0a=10,\tau\varepsilon_0/\hbar=10, h/\varepsilon_0=0.1, \alpha q_0^3/\varepsilon_0=0.2$, and $\mu/\varepsilon_0=0$. Here, $\varepsilon_0$ and $q_0=\sqrt{2m\varepsilon_0}/\hbar$ are arbitrary energy and length scales, respectively. $\mu$ is the chemical potential. }
        \label{fig:2DSK}
    \end{center}
\end{figure}

\textit{2.~Momentum space}. Next, we consider the case where the $\bm{r}$-dependence of the system appears only through the external potential, such that the wavefunction depends solely on $\bm{q}$. By removing the terms involving derivatives with respect to $\lambda$ and $t$ from the Lagrangian, the resulting equation of motion becomes
\begin{align}
    \dot{q}_a&=-\frac{1}{\hbar}\frac{\partial V}{\partial r_a},\\
    \dot{r}_a&=\frac{1}{\hbar}\frac{\partial \varepsilon_n}{\partial q_a}+\frac{1}{\hbar}\Omega^{qq}_{ai}\frac{\partial V}{\partial r_i}+\frac{1}{\hbar}\Gamma^{qqq}_{a,ij}\frac{\partial V}{\partial r_i}\frac{\partial V}{\partial r_j}.
\end{align}
Then, the current in the lowest order of $\tau$ is given by 
\begin{align}
    j^{(0)}_a&=-\frac{e}{\hbar}\int \left(\frac{\partial \varepsilon_n}{\partial q_a}+e\Omega^{qq}_{ai}E_i+e^2\Gamma^{qqq}_{a,ij}E_iE_j\right)f_0.
\end{align}
The term involving $\Omega^{qq}$ corresponds to the conventional anomalous Hall effect.
The term with $\Gamma^{qqq}$ contains both symmetric and antisymmetric components under the exchange of the subindices $a$ and $i$ or $j$. 
The nonlinear Hall conductivity is thus given by 
\begin{align}
    \sigma^{\mathrm{nl}}_{ab}=\frac{-e^3}{2\hbar}\int \left(\Gamma^{qqq}_{a,bb}-\Gamma^{qqq}_{b,aa}\right)f_0.
\end{align}
We see that the antisymmetric part of $\Gamma^{qqq}$ gives rise to a nonlinear Hall effect induced by momentum-space geometry which we refer to as "momentum-gravity" in the following. 

\begin{figure}[t]
    \begin{center}
        \includegraphics[width=\columnwidth]{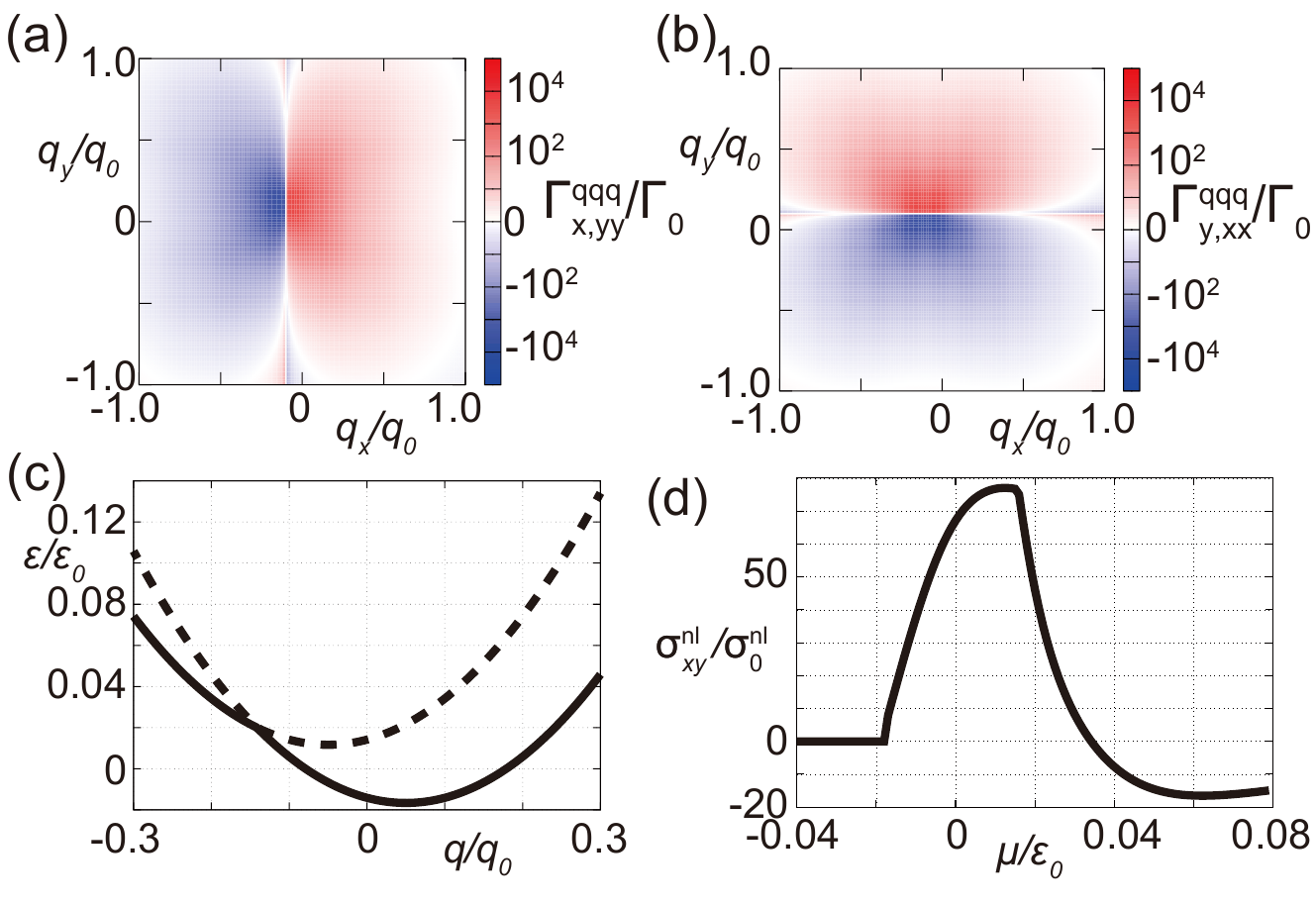}
        \caption{Momentum-gravity in a system with Rashba-type spin-orbit interaction and a uniform magnetic field. Christoffel symbols (a) $\Gamma^{qqq}_{x,yy}$ and (b) $\Gamma^{qqq}_{y,xx}$ in momentum space are plotted. (c) The band structure of the system. (d) Nonlinear Hall conductivity as a function of chemical potential at zero temperature. Parameters are $C q_0/\varepsilon_0=0.1, H_x/\varepsilon_0=0.01,$ and $H_y/\varepsilon_0=0.01$, where $\varepsilon_0$ and $q_0=\sqrt{2m\varepsilon_0}/\hbar$ are arbitrary energy and length scales, respectively. Here, $\Gamma_0=1/(\varepsilon_0q_0^3)$ and $\sigma^{\mathrm{nl}}_0=e^3/(\hbar\varepsilon_0q_0)$.}
        \label{fig:Rashba}
    \end{center}
\end{figure}

As an example, we consider a spatially uniform two-dimensional system with Rashba-type spin-orbit interaction and Zeeman coupling. The Hamiltonian is given by
\begin{align}
    H(\bm{q})&=\frac{\hbar^2|\bm{q}|^2}{2m}+C(\hat{z}\times\bm{q})\cdot\boldsymbol{\sigma}+\bm{H}\cdot\boldsymbol{\sigma}\label{eq:Rashba_Hamil},
\end{align}
where $\hat{z}$ is a unit vector parallel to the $z$ axis, $C$ is a constant, and $\bm{H}=\left(H_x,H_y,0\right)^{T}$ is an external magnetic field. For this model, $\Gamma^{qqq}_{x,yy}$ and $\Gamma^{qqq}_{y,xx}$ are shown in Figs.~\ref{fig:Rashba}~(a) and (b), respectively (see Supplemental Material~\cite{supplementary} for the expressions of the Christoffel symbols). The difference of these two quantities contributes to $\sigma^{\mathrm{nl}}_{xy}$. As we can see from the band structure in Fig.~\ref{fig:Rashba}~(c), the Christoffel symbol diverges at the gap-closing point (see Supplemental Material~\cite{supplementary}). However, near the gap-closing point, the perturbation theory used to derive the Lagrangian is no longer valid and this divergence would not occur realistically. Still, we expect a finite nonlinear Hall conductivity even when the chemical potential is far from gap-closing point as shown in Fig.~\ref{fig:Rashba}~(d). Realistically, the Zeeman coupling is on the order of meV, leading to $\varepsilon_0\approx0.1$ eV. In this case, $\sigma^{\mathrm{nl}}_{xy}\sim10^{-9}$ to $10^{-10}$ Am/V$^2$, which is larger than the nonlinear Hall effect induced by the Berry curvature dipole~\cite{sodemann_quantum_2015}, typically on the order of $10^{-12}$ Am/V$^2$. While the latter scales as $\tau^1$, the momentum-gravity-induced nonlinear Hall effect scales as $\tau^0$, allowing the two mechanisms to be distinguished. The in-plane magnetic field also induces a nonlinear planar Hall effect~\cite{He2019,Esin2021,Huang_2023}. Its intrinsic contribution shows the same scaling, $\tau^0$,~\cite{Huang_2023} and is estimated to be on the order of $10^{-12}$ Am/V$^2$, which is significantly smaller than that from the momentum-gravity Hall effect.

Finally, we comment on symmetry constraints. By construction, the intrinsic second-order Hall response requires the breaking of both time-reversal and inversion symmetries. Furthermore, the antisymmetric part of $\sigma^{\mathrm{nl}}_{ab}$ transforms under crystallographic point groups in the same way as that for the nonlinear Hall effect in Ref.~\cite{duQuantumTheoryNonlinear2021a}.

\textit{3.~Time domain}. 
We now focus on the time dependence of the system.  

We examine the role of Christoffel symbols in mixed real, momentum, and time spaces. Neglecting the $\lambda$ dependence, the terms involving Christoffel symbols with time derivatives in Eqs.~\eqref{eq:General_EOM_q} and \eqref{eq:General_EOM_r} are given by
\begin{align}
    \dot{q}_a =&(-2\Gamma^{rrt}_{a,i\phantom{j}}\dot{r}_i-2\Gamma^{rqt}_{a,i\phantom{j}}\dot{q}_i-\Gamma^{rtt}_{a,})\hbar+\cdots,\label{eq:t-EOM_q}\\
    \dot{r}_a =&(2\Gamma^{qrt}_{a,i\phantom{j}}\dot{r}_i+2\Gamma^{qqt}_{a,i\phantom{j}}\dot{q}_i+\Gamma^{qtt}_{a,})\hbar+\cdots\label{eq:t-EOM_r}.
\end{align}
The physical meaning of each term can be understood by analogy.  
The antisymmetric part of $\Gamma^{qqt}$ induces a velocity perpendicular to $\dot{\bm{q}}$  analogous to the anomalous velocity associated with the momentum-space Berry curvature $\Omega^{qq}$.
Thus, this gravitational anomalous velocity results in a Hall effect. 
Similarly, the antisymmetric part of $\Gamma^{rrt}$ gives rise to a force  perpendicular to $\dot{\bm{r}}$ and acts like a magnetic field in real space. 
Hence, this gravitational Lorentz force also leads to a Hall effect. 

The mixed-space terms $\Gamma^{rqt}$ and $\Gamma^{qrt}$ resemble the terms $\Omega^{rq}$ and $\Omega^{qr}$, respectively, and give modifications of density of states from mixed-space geometry. In contrast, $\Gamma^{rtt}$ and $\Gamma^{qtt}$ behave as external forces and group velocities due to real-space and momentum-space gravity, respectively.

As an example, reconsider the magnetic skyrmion discussed above. If the center of the skyrmion moves with a constant velocity $v$ along the $x$ axis, its position is given by $vt$. Thus, we replace $x$ by $x-vt$ in Eq.~\eqref{eq_sky_u}.  
In this case, spatial and temporal derivatives of the Christoffel symbols are related by a constant factor. For example, we have $\Gamma^{rrt}_{y,x}=-v\Gamma^{rrr}_{y,xx}$. Therefore, we see that for a magnetic skyrmion, a Hall effect due to the gravitational Lorentz force arises. Also, for $\Gamma^{qqt}$, reconsider the system described by Eq.~\eqref{eq:Rashba_Hamil}. Suppose that the magnetic field changes in time $\bm{H}(t)$. Then, we have $\partial_t=\dot{H}_i\partial_{H_i}$ and since magnetic field only shifts the momentum in Eq.~\eqref{eq:Rashba_Hamil}, $\partial_{H_i}=\varepsilon_{izj}/C\partial_{q_j}$ holds. Therefore, we obtain $\Gamma^{qqt}_{a,b}=\dot{H}_{i}/C\varepsilon_{izj}\Gamma^{qqq}_{a,bj}$ and find that this gravitational anomalous velocity also causes a Hall effect.

According to Eqs.~\eqref{eq:General_EOM_q} and \eqref{eq:General_EOM_r}, charge current can be induced by dynamics of an external parameter $\lambda$.
We discuss charge pumping by parameter space gravity in Supplemental Material~\cite{supplementary}.

\textit{Conclusion}. 
We have investigated transport phenomena in systems with spatial, momentum, temporal, and external-parameter dependencies. By incorporating nonadiabatic effects into the semiclassical description of electron wave packets, we find that various Christoffel symbols govern the dynamics and hence transport phenomena. We propose the emergent-gravity Hall effect, driven by real-space gravity encoded in  the real-space Christoffel symbol, which appears in systems such as a single magnetic skyrmion.
We have also shown that Christoffel symbols in momentum space lead to nonlinear Hall responses, while those involving time dependencies describe gravitational anomalous velocity and gravitational Lorentz force, also giving rise to Hall effects. Our formulation treats all variables on an equal footing and offers a unified framework for investigating effects of effective gravity in quantum systems.

\textit{Note added}. During the preparation of our manuscript, we became aware of two related works. Reference~\cite{onishi_emergent_2025} discusses emergent gravity from real-space spin textures based on a U(2) gauge field. Reference~\cite{ren_momentum-space_2025} develops a nonadiabatic formalism similar to this paper, but limited to momentum-space effects. 
These works do not discuss Hall effects. During the review process, Ref.~\cite{ren_analogue_2025} appeared, which discusses emergent gravity in a general phase-space framework but does not address transport effects.

\textit{Acknowledgements}. HY is supported by Japan Society for the Promotion of Science (JSPS) KAKENHI Grant Number JP24KJ1109 and by MEXT Initiative to Establish Next-generation Novel Integrated Circuits Centers (X-NICS) Grant Number JPJ011438. TY is supported by Japan Society for the Promotion of Science (JSPS) KAKENHI Grant Numbers 25K07221.

\bibliography{EG.bib}

\clearpage
\newpage

\renewcommand{\theequation}{S\arabic{equation}}
\setcounter{equation}{1}
\setcounter{page}{1}

% %%%%%%%%%%%%%%%%%%%%%%%%%%%%%%%%%%%%%%%%%%%%%%%%%%%%%%%%%%%%%%%%%%%%%%%%%%%%%%

% \begin{document}

\onecolumngrid

\begin{center}
{\Large\bfseries Emergent-gravity Hall effect from quantum geometry: Supplementary Material}\\[1em]
{\normalsize Hiroki Yoshida and Takehito Yokoyama}\\[0.5em]
{\small Department of Physics, Institute of Science Tokyo, 2-12-1 Ookayama, Meguro-ku, Tokyo 152-8551, Japan}
\end{center}

% \title{Emergent-gravity Hall effect from quantum geometry: Supplementary Material}

% \author{Hiroki Yoshida}
% \affiliation{Department of Physics, Institute of Science Tokyo, 2-12-1 Ookayama, Meguro-ku, Tokyo 152-8551, Japan}
% \author{Takehito Yokoyama}
% \affiliation{Department of Physics, Institute of Science Tokyo, 2-12-1 Ookayama, Meguro-ku, Tokyo 152-8551, Japan}

% \maketitle

% \tableofcontents

\makeatletter

% Supplementary専用ToCを表示
\section*{CONTENTS}
\@starttoc{stoc}

% 以後、Supplementary内のsection情報を .toc ではなく .stoc に書く
\let\mainaddcontentsline\addcontentsline
\renewcommand{\addcontentsline}[3]{%
  \def\tempa{#1}%
  \def\tempb{toc}%
  \ifx\tempa\tempb
    \mainaddcontentsline{stoc}{#2}{#3}%
  \else
    \mainaddcontentsline{#1}{#2}{#3}%
  \fi
}

\makeatother

\section{Derivation of the Lagrangian}

In this section, we derive the Lagrangian used in the main text, keeping the leading nonadiabatic correction to the wavepacket state. We start from the time-dependent Schr\"odinger equation
\begin{equation}
    i\hbar\frac{\mathrm d}{\mathrm dt}\lvert\Psi\rangle=\hat{H}\lvert\Psi\rangle .
\end{equation}
The electron wave packet is written as
\begin{equation}
    \lvert\Psi\rangle\coloneqq \int \mathrm d\bm q\, a(\bm q,t)\lvert\psi_{\bm q}(\bm r_c,t)\rangle,
\end{equation}
where it is assumed to be sharply peaked at the wavepacket center in momentum space,
\begin{equation}
    \lvert a(\bm q,t)\rvert^2\approx \delta(\bm q-\bm q_c).
\end{equation}
The real-space center is defined by
\begin{equation}
    \bm r_c=\langle\Psi\rvert \hat{\bm r}\lvert\Psi\rangle .
\end{equation}
We use the Bloch-state convention
\begin{equation}
    \lvert\psi_{\bm q}(\bm r_c,t)\rangle=e^{i\bm q\cdot \hat{\bm r}}\lvert u_{\bm q}(\bm r_c,t)\rangle,
\end{equation}
where \(\lvert u_{\bm q}(\bm r_c,t)\rangle\) is the periodic part of the Bloch state. With this convention the wavepacket Lagrangian of Ref.~\cite{sundaram_wave-packet_1999} is
\begin{align}
    L=&-\langle\Psi\rvert\hat{H}\lvert\Psi\rangle+
    \hbar\bm q_c\cdot\dot{\bm r}_c
    +\hbar\dot{\bm q}_c\cdot\left\langle u\middle|i\frac{\partial u}{\partial\bm q_c}\right\rangle
    +\hbar\dot{\bm r}_c\cdot\left\langle u\middle|i\frac{\partial u}{\partial\bm r_c}\right\rangle
    +\hbar\left\langle u\middle|i\frac{\partial u}{\partial t}\right\rangle .
    \label{eq:Lagrangian}
\end{align}
The first-order canonical term may equivalently be written as \(-\hbar\dot{\bm q}_c\cdot\bm r_c\), since the two forms differ only by the total time derivative \(\mathrm d(\hbar\bm q_c\cdot\bm r_c)/\mathrm dt\).

Reference~\cite{sundaram_wave-packet_1999} considered a single band in the adiabatic approximation. Here we include the leading nonadiabatic mixing with other instantaneous bands. The instantaneous Bloch wave eigenstates are defined by
\begin{equation}
    \hat H(\bm q,\bm r_c,t)\lvert\psi_{\alpha,\bm q}(\bm r_c,t)\rangle
    =\varepsilon_\alpha(\bm q,\bm r_c,t)\lvert\psi_{\alpha,\bm q}(\bm r_c,t)\rangle .
\end{equation}
For a fixed wave vector, the time-dependent Bloch state satisfies
\begin{equation}
    i\hbar\frac{\mathrm d}{\mathrm dt}\lvert\psi_{\bm q}(\bm r_c,t)\rangle
    =\hat H(\bm q,\bm r_c,t)\lvert\psi_{\bm q}(\bm r_c,t)\rangle .
    \label{eq:TDS_Bloch}
\end{equation}
Following the standard adiabatic expansion, we write
\begin{equation}
    \lvert\psi_{\bm q}(\bm r_c,t)\rangle
    =\sum_\alpha \exp\left[-\frac{i}{\hbar}\int_{t_0}^{t}\mathrm dt'\,\varepsilon_\alpha(t')\right]
    c_\alpha(t)\lvert\psi_{\alpha,\bm q}(\bm r_c,t)\rangle .
\end{equation}
Substitution into Eq.~\eqref{eq:TDS_Bloch} gives
\begin{equation}
    \dot c_\alpha(t)=
    -\sum_\beta \exp\left[-\frac{i}{\hbar}\int_{t_0}^{t}\mathrm dt'\,
    \{\varepsilon_\beta(t')-\varepsilon_\alpha(t')\}\right]
    \left\langle\psi_{\alpha,\bm q}(t)\middle|\frac{\mathrm d}{\mathrm dt}\psi_{\beta,\bm q}(t)\right\rangle c_\beta(t).
    \label{eq:coefficient_equation}
\end{equation}
We assume that the state is initially in the reference band \(n\), namely \(c_\alpha(t_0)=\delta_{\alpha n}\). For \(m\neq n\), the first-order term obtained by iterating Eq.~\eqref{eq:coefficient_equation} is
\begin{align}
    I_{mn}(t)=&\int_{t_0}^{t}\mathrm dt'\,
    \exp\left[-\frac{i}{\hbar}\int_{t_0}^{t'}\mathrm dt''\,
    \{\varepsilon_n(t'')-\varepsilon_m(t'')\}\right]
    \left\langle\psi_{m,\bm q}(t')\middle|\frac{\mathrm d}{\mathrm dt'}\psi_{n,\bm q}(t')\right\rangle .
\end{align}
If the matrix element multiplying the oscillatory phase varies slowly on the time scale set by the band separation, integration by parts gives
\begin{align}
    I_{mn}(t)
    &\approx\left\{
    \frac{i\hbar\left\langle\psi_{m,\bm q}(t)\middle|\frac{\mathrm d}{\mathrm dt}\psi_{n,\bm q}(t)\right\rangle}
    {\varepsilon_n(t)-\varepsilon_m(t)}
    -\frac{i\hbar}{\varepsilon_n(t)-\varepsilon_m(t)}\frac{\rmd}{\rmd t}\left(\frac{i\hbar\left\langle\psi_{m,\bm q}(t)\middle|\frac{\mathrm d}{\mathrm dt}\psi_{n,\bm q}(t)\right\rangle}
    {\varepsilon_n(t)-\varepsilon_m(t)}\right)
    \right\}\exp\left[-\frac{i}{\hbar}\int_{t_0}^{t}\mathrm dt'\,
    \{\varepsilon_n(t')-\varepsilon_m(t')\}\right] .
    \label{eq:first_order_integral}
\end{align}
The boundary term at \(t=t_0\) is neglected by assuming that the initial switching-on is adiabatic. The further derivatives of the prefactor in Eq.~\eqref{eq:first_order_integral} give higher-order nonadiabatic corrections and are dropped here.

The second-order term is obtained in the same way. For \(m\neq n\),
\begin{align}
    I'_{mn}(t)=&\int_{t_0}^{t}\mathrm dt'\int_{t_0}^{t'}\mathrm dt''
    \sum_{l\neq n}
    \exp\left[-\frac{i}{\hbar}\int_{t_0}^{t'}\mathrm ds\,
    \{\varepsilon_l(s)-\varepsilon_m(s)\}\right]
    \left\langle\psi_{m}(t')\middle|\frac{\mathrm d}{\mathrm dt'}\psi_l(t')\right\rangle
    \nonumber\\
    &\hspace{2.0cm}\times
    \exp\left[-\frac{i}{\hbar}\int_{t_0}^{t''}\mathrm ds\,
    \{\varepsilon_n(s)-\varepsilon_l(s)\}\right]
    \left\langle\psi_l(t'')\middle|\frac{\mathrm d}{\mathrm dt''}\psi_n(t'')\right\rangle
    \nonumber\\
    &\approx
    \sum_{l\neq n}
    \frac{i\hbar\left\langle\psi_l(t)\middle|\frac{\mathrm d}{\mathrm dt}\psi_n(t)\right\rangle}
    {\varepsilon_n(t)-\varepsilon_l(t)}
    \frac{i\hbar\left\langle\psi_m(t)\middle|\frac{\mathrm d}{\mathrm dt}\psi_l(t)\right\rangle}
    {\varepsilon_n(t)-\varepsilon_m(t)}
    \exp\left[-\frac{i}{\hbar}\int_{t_0}^{t}\mathrm dt'\,
    \{\varepsilon_n(t')-\varepsilon_m(t')\}\right] .
    \label{eq:second_order_integral}
\end{align}
Here and below, diagonal terms in the total time derivative may be absorbed into the phase convention for the instantaneous eigenstates.

Introducing
\begin{equation}
    A_{\alpha\beta}\coloneqq
    \left\langle\psi_\alpha\middle|i\frac{\mathrm d}{\mathrm dt}\psi_\beta\right\rangle,
\end{equation}
and factoring out the common dynamical phase of the reference band, the wave function up to \(\mathcal O(\hbar^2)\) is
\begin{align}
    \lvert\psi_{\bm q}(\bm r_c,t)\rangle=&
    \exp\left[-\frac{i}{\hbar}\int_{t_0}^{t}\mathrm dt'\,\varepsilon_n(t')\right]
    \Bigg[
    \left(1-\frac{\hbar^2}{2}\sum_{m\neq n}
    \frac{A_{mn}A_{nm}}{(\varepsilon_{n,c}-\varepsilon_{m,c})^2}\right)
    \lvert\psi_{n,\bm q}(\bm r_c,t)\rangle
    \nonumber\\
    &+
    \sum_{m\neq n}
    \left(-\hbar\frac{A_{mn}}{\varepsilon_{n,c}-\varepsilon_{m,c}}
    +\hbar^2\sum_{l\neq n}
    \frac{A_{ml}A_{ln}}
    {(\varepsilon_{n,c}-\varepsilon_{l,c})(\varepsilon_{n,c}-\varepsilon_{m,c})}+i\hbar^2\frac{1}{\varepsilon_{n,c}-\varepsilon_{m,c}}\frac{\rmd}{\rmd t}\left(\frac{A_{mn}}{\varepsilon_{n,c}-\varepsilon_{m,c}}\right)\right)
    \lvert\psi_{m,\bm q}(\bm r_c,t)\rangle
    \Bigg] .
\end{align}
The coefficient of the reference band is fixed by normalization, as in the ordinary perturbation theory. Equivalently, for the periodic part entering the wavepacket Lagrangian, we obtain
\begin{align}
    \lvert u\rangle=&
    \left(1-\frac{\hbar^2}{2}\sum_{m\neq n}
    \frac{A_{mn}A_{nm}}{(\varepsilon_{n,c}-\varepsilon_{m,c})^2}\right)
    \lvert u_n(t)\rangle\nonumber\\
    &+\sum_{m\neq n}
    \left(-\hbar\frac{A_{mn}}{\varepsilon_{n,c}-\varepsilon_{m,c}}
    +\hbar^2\sum_{l\neq n}
    \frac{A_{ml}A_{ln}}
    {(\varepsilon_{n,c}-\varepsilon_{l,c})(\varepsilon_{n,c}-\varepsilon_{m,c})}+i\hbar^2\frac{1}{\varepsilon_{n,c}-\varepsilon_{m,c}}\frac{\rmd}{\rmd t}\left(\frac{A_{mn}}{\varepsilon_{n,c}-\varepsilon_{m,c}}\right)\right)
    \lvert u_m(t)\rangle .
    \label{eq:Bloch_hbar2}
\end{align}

We next insert Eq.~\eqref{eq:Bloch_hbar2} into Eq.~\eqref{eq:Lagrangian}. Let \(X\) denote any generalized coordinate on which the instantaneous eigenstates depend, such as \(\bm q\), \(\bm r\), an additional parameter \(\bm\lambda\), or time \(t\) itself. For \(X=t\), the component index is omitted and \(\dot t=1\). We define
\begin{equation}
    A^X_{\alpha\beta,i}\coloneqq
    \left\langle u_\alpha\middle|i\frac{\partial u_\beta}{\partial X_i}\right\rangle,
    \qquad
    A_{\alpha\beta}=\sum_X \dot X_i A^X_{\alpha\beta,i} .
\end{equation}
Only the \(\mathcal O(\hbar)\) part of \(\lvert u\rangle\) is required in the Berry-connection terms of Eq.~\eqref{eq:Lagrangian}. Writing \(\lvert u\rangle=\lvert u_n\rangle+\sum_{m\neq n}C_m\lvert u_m\rangle+\mathcal O(\hbar^2)\), with
\begin{equation}
    C_m=\hbar\sum_Y\frac{\dot Y_i A^Y_{mn,i}}{\varepsilon_m-\varepsilon_n},
\end{equation}
we find
\begin{align}
    \left\langle u\middle|i\frac{\partial u}{\partial X_j}\right\rangle
    &=A^X_{nn,j}+2\sum_{m\neq n}\operatorname{Re}\left[C_m A^X_{nm,j}\right]+\mathcal O(\hbar^2)
    \nonumber\\
    &=A^X_{nn,j}+\hbar\sum_Y G^{XY}_{ji}\dot Y_i+\mathcal O(\hbar^2),
\end{align}
where the weighted quantum metric of the reference band is
\begin{equation}
    G^{XY}_{ij}
    =2\operatorname{Re}\left[\sum_{m\neq n}
    \frac{A^X_{nm,i}A^Y_{mn,j}}{\varepsilon_m-\varepsilon_n}\right] .
    \label{eq:weighted_metric}
\end{equation}

The energy term requires \(\lvert u\rangle\) up to \(\mathcal O(\hbar^2)\). A direct calculation gives
\begin{align}
    \langle\Psi\rvert\hat H\lvert\Psi\rangle
    &=\varepsilon_{n,c}-\hbar^2\sum_{m\neq n}
    \frac{A_{nm}A_{mn}}{\varepsilon_{n,c}-\varepsilon_{m,c}}
    \nonumber\\
    &=\varepsilon_{n,c}+\frac{\hbar^2}{2}\sum_{X,Y}\dot X_iG^{XY}_{ij}\dot Y_j .
\end{align}
Combining this energy correction with the Berry-connection correction yields the effective Lagrangian for a wave packet initially prepared in band \(n\):
\begin{equation}
    L=-\varepsilon_{n,c}+\hbar\bm q_c\cdot\dot{\bm r}_c
    +\hbar\sum_X\dot X_iA^X_{nn,i}
    +\frac{\hbar^2}{2}\sum_{X,Y}\dot X_iG^{XY}_{ij}\dot Y_j .
    \label{eq:Lagrangian_derived}
\end{equation}
In the following we suppress the center subscript \(c\) unless it is needed explicitly.

\section{Derivation of the Equations of Motion}

Starting from Eq.~\eqref{eq:Lagrangian_derived}, we derive the equations of motion by the Euler--Lagrange equations. We use
\begin{equation}
    \Omega^{XY}_{ij}\coloneqq
    \frac{\partial A^Y_{nn,j}}{\partial X_i}
    -\frac{\partial A^X_{nn,i}}{\partial Y_j}
\end{equation}
and
\begin{equation}
    \Gamma^{XYZ}_{aij}
    \coloneqq \frac{1}{2}\left(
    \frac{\partial G^{XZ}_{aj}}{\partial Y_i}
    +\frac{\partial G^{XY}_{ai}}{\partial Z_j}
    -\frac{\partial G^{YZ}_{ij}}{\partial X_a}
    \right) .
    \label{eq:christoffel_definition}
\end{equation}
For the equation conjugate to \(q_a\),
\begin{align}
    \frac{\mathrm d}{\mathrm dt}\left(\frac{\partial L_n}{\partial\dot q_a}\right)
    &=
    \hbar\sum_X\frac{\partial A^q_{nn,a}}{\partial X_i}\dot X_i
    +\hbar^2\sum_X\left(G^{qX}_{ai}\ddot X_i
    +\sum_Y\dot X_i\frac{\partial G^{qX}_{ai}}{\partial Y_j}\dot Y_j\right),
    \\
    \frac{\partial L_n}{\partial q_a}
    &=-\frac{\partial\varepsilon_n}{\partial q_a}
    +\hbar\dot r_a
    +\hbar\sum_X\frac{\partial A^X_{nn,i}}{\partial q_a}\dot X_i
    +\frac{\hbar^2}{2}\sum_{X,Y}\dot X_i
    \frac{\partial G^{XY}_{ij}}{\partial q_a}\dot Y_j .
\end{align}
Equating these expressions gives
\begin{align}
    \dot r_a
    &=\frac{1}{\hbar}\frac{\partial\varepsilon_n}{\partial q_a}
    -\sum_X\Omega^{qX}_{ai}\dot X_i
    +\hbar\sum_{X,Y}\Gamma^{qXY}_{aij}\dot X_i\dot Y_j
    +\hbar\sum_XG^{qX}_{ai}\ddot X_i .
    \label{eq:eom_r_general}
\end{align}
Similarly, the equation conjugate to \(r_a\) is obtained from
\begin{align}
    \frac{\mathrm d}{\mathrm dt}\left(\frac{\partial L_n}{\partial\dot r_a}\right)
    &=
    \hbar\dot q_a
    +\hbar\sum_X\frac{\partial A^r_{nn,a}}{\partial X_i}\dot X_i
    +\hbar^2\sum_X\left(G^{rX}_{ai}\ddot X_i
    +\sum_Y\dot X_i\frac{\partial G^{rX}_{ai}}{\partial Y_j}\dot Y_j\right),
    \\
    \frac{\partial L_n}{\partial r_a}
    &=-\frac{\partial\varepsilon_n}{\partial r_a}
    +\hbar\sum_X\frac{\partial A^X_{nn,i}}{\partial r_a}\dot X_i
    +\frac{\hbar^2}{2}\sum_{X,Y}\dot X_i
    \frac{\partial G^{XY}_{ij}}{\partial r_a}\dot Y_j .
\end{align}
Thus, we obtain
\begin{equation}
    \hbar\dot q_a
    =-\frac{\partial\varepsilon_n}{\partial r_a}
    +\hbar\sum_X\Omega^{rX}_{ai}\dot X_i
    -\hbar^2\sum_{X,Y}\Gamma^{rXY}_{aij}\dot X_i\dot Y_j
    -\hbar^2\sum_XG^{rX}_{ai}\ddot X_i .
    \label{eq:eom_q_general}
\end{equation}

\section{Emergent-Gravity Hall Effect}

We now include an external perturbation \(\hat H'\). As in the conventional semiclassical theory through \(\mathcal O(\hbar)\), the perturbation modifies the band energy and the instantaneous Bloch states through the ordinary perturbation theory. For a uniform electric field, the Hamiltonian is
\begin{align}
    \hat H
    &=\hat H_0+e\bm E\cdot\hat{\bm r}
    \nonumber\\
    &=\hat H_0+e\bm E\cdot\bm r_c+e\bm E\cdot(\hat{\bm r}-\bm r_c),
\end{align}
where \(e>0\) denotes the elementary charge, so that the electron charge is \(-e\). The wavepacket energy is
\begin{equation}
    \tilde\varepsilon
    =\langle\tilde\Psi\rvert\hat H\lvert\tilde\Psi\rangle
    =\varepsilon+e\bm E\cdot\bm r_c+\tilde\varepsilon^{(1)}_n+\tilde\varepsilon^{(2)}_n+\cdots .
\end{equation}
Here \(\lvert\tilde\Psi\rangle\) is a wave packet constructed from the perturbed states,
\begin{equation}
    \varepsilon=\varepsilon_n+\frac{\hbar^2}{2}\sum_{X,Y}\dot X_iG^{XY}_{ij}\dot Y_j
\end{equation}
is the unperturbed wavepacket energy including the nonadiabatic correction, and \(\tilde\varepsilon^{(m)}_n\) is the \(m\)th-order correction due to \(e\bm E\cdot(\hat{\bm r}-\bm r_c)\). The first-order correction from this last term vanishes by the definition of the wavepacket center.

\subsection{Real-space texture}

First, we consider the case in which the periodic part of the Bloch state depends on real space but has no Berry connection in momentum space. Introducing \(p_a=\hbar q_a\), and using the equivalent canonical form \(\bm p\cdot\dot{\bm r}\approx-\dot{\bm p}\cdot\bm r\), the Lagrangian through \(\mathcal O(\hbar^2)\) can be written as
\begin{align}
    L
    &=-\tilde\varepsilon-\dot{\bm p}\cdot\bm r
    +\hbar A^r_{nn,j}\dot r_j
    +\hbar^2G^{rr}_{ij}\dot r_i\dot r_j
    \nonumber\\
    &=-\varepsilon_n-e\bm E\cdot\bm r-\dot{\bm p}\cdot\bm r
    +\hbar A^r_{nn,j}\dot r_j
    +\frac{\hbar^2}{2}G^{rr}_{ij}\dot r_i\dot r_j.
    \label{eq:Lagrangian_real}
\end{align}
The corresponding equations of motion are
\begin{equation}
    \begin{cases}
        \dot r_a=v_a,\\[2mm]
        \dot p_a=-eE_a+
        \hbar\Omega^{rr}_{ai}\dot r_i
        -\hbar^2\Gamma^{rrr}_{aij}\dot r_i\dot r_j
        -\hbar^2G^{rr}_{aj}\ddot r_j,
    \end{cases}
    \label{eq:eom_real}
\end{equation}
where
\begin{equation}
    v_a\coloneqq \frac{\partial\varepsilon_n}{\partial p_a} .
\end{equation}
Up to \(\mathcal O(\hbar^2)\) terms, \(\dot r_a\) can be replaced by \(v_a\). The unperturbed energy entering the equilibrium distribution is therefore
\begin{equation}
    \varepsilon
    =\varepsilon_n(\bm p)+\frac{\hbar^2}{2}G^{rr}_{ij}v_i v_j .
    \label{eq:real_energy_correction}
\end{equation}
For a fixed band \(n\), the current to linear order in the electric field is given by
\begin{equation}
    j_a=-e\int\frac{\mathrm d^3p}{(2\pi\hbar)^3}\,D\,f(\varepsilon)\,v_a .
    \label{eq:R-EGHE_current}
\end{equation}
A band sum should be added to Eq.~\eqref{eq:R-EGHE_current} when several bands cross the Fermi level.

\subsubsection{Density of states}

The density-of-states factor \(D\) contains the phase-space-volume correction induced by the quadratic term in the Lagrangian as pointed out in Refs.~\cite{MamedaYamamoto,maranzana_shinada_nagaosa2026}. For Eq.~\eqref{eq:Lagrangian_real}, the correction through \(\mathcal O(\hbar^2)\) is
\begin{equation}
    D=1+\hbar^2G^{rr}_{ij}\frac{\partial^2\varepsilon_n}{\partial p_i\partial p_j} .
    \label{eq:dos_real}
\end{equation}

\subsubsection{Distribution function}

The distribution function is obtained from the Boltzmann equation in the relaxation-time approximation,
\begin{equation}
    \dot{\bm r}\cdot\frac{\partial f}{\partial\bm r}
    +\dot{\bm p}\cdot\frac{\partial f}{\partial\bm p}
    =-\frac{f-f_0}{\tau},
\end{equation}
where \(f_0\) is the equilibrium distribution function and \(\tau\) is the relaxation time. Because of Eq.~\eqref{eq:real_energy_correction}, \(f_0\) depends on both \(\bm r\) and \(\bm p\) through \(\varepsilon\). We assume \(l/L\ll1\), where \(l\) is the mean free path and \(L\) is the characteristic length scale of the real-space texture. Terms higher order in this gradient expansion are not retained.

Defining a differential operator
\begin{equation}
    \frac{\rmd}{\rmd t}= \dot r_a\frac{\partial}{\partial r_a}
    +\dot p_a\frac{\partial}{\partial p_a},
\end{equation}
the iterative solution through \(\tau^2\) is
\begin{align}
    f
    &\approx f_0(\varepsilon)-\tau\frac{\rmd}{\rmd t} f_0(\varepsilon)
    +\tau^2\frac{\rmd}{\rmd t}^2 f_0(\varepsilon)
    \nonumber\\
    &=f_0
    -\tau\frac{\partial f_0}{\partial\varepsilon}\frac{\rmd}{\rmd t}\varepsilon
    +\tau^2\left[
    \frac{\partial^2f_0}{\partial\varepsilon^2}\left(\frac{\rmd}{\rmd t}\varepsilon\right)^2
    +\frac{\partial f_0}{\partial\varepsilon}\frac{\rmd^2}{\rmd t^2}\varepsilon
    \right] .
\end{align}
Keeping terms through \(\mathcal O(\hbar^2)\) and linear order in \(\bm E\), we obtain
\begin{align}
    \frac{\rmd}{\rmd t}\varepsilon
    &\approx -eE_i v_i,\\
    \left(\frac{\rmd}{\rmd t}\varepsilon\right)^2
    &\approx 0,\\
    \frac{\rmd^2}{\rmd t^2}\varepsilon
    &\approx
    -eE_a\left(\hbar\Omega^{rr}_{bj}
    -\hbar^2\Gamma^{rrr}_{bjk}v_k\right)
    v_j\frac{\partial^2\varepsilon_n}{\partial p_a\partial p_b},\\
    f_0(\varepsilon)
    &\approx f_0(\varepsilon_n)
    +\frac{\hbar^2}{2}G^{rr}_{ij}v_i v_j
    \frac{\partial f_0(\varepsilon_n)}{\partial\varepsilon_n},\\
    \frac{\partial f_0(\varepsilon)}{\partial\varepsilon}
    &\approx
    \frac{\partial f_0(\varepsilon_n)}{\partial\varepsilon_n}
    +\frac{\hbar^2}{2}G^{rr}_{ij}v_i v_j
    \frac{\partial^2 f_0(\varepsilon_n)}{\partial\varepsilon_n^2} .
\end{align}
Thus, we obtain the expression of the distribution function: 
\begin{align}
    f(\varepsilon)
    =& f_0(\varepsilon_n)
    +\frac{\hbar^2}{2}G^{rr}_{ij}v_i v_j
    \frac{\partial f_0(\varepsilon_n)}{\partial\varepsilon_n}
    \nonumber\\
    &+\tau eE_a v_a\left[
    \frac{\partial f_0(\varepsilon_n)}{\partial\varepsilon_n}
    +\frac{\hbar^2}{2}G^{rr}_{ij}v_i v_j
    \frac{\partial^2 f_0(\varepsilon_n)}{\partial\varepsilon_n^2}
    \right]
    \nonumber\\
    &-\tau^2 eE_a\left(\hbar\Omega^{rr}_{bj}
    -\hbar^2\Gamma^{rrr}_{bjk}v_k\right)
    v_j\frac{\partial^2\varepsilon_n}{\partial p_a\partial p_b}
    \frac{\partial f_0(\varepsilon_n)}{\partial\varepsilon_n} .
    \label{eq:distribution_real}
\end{align}

\subsubsection{Electric current}

Using Eqs.~\eqref{eq:R-EGHE_current}, \eqref{eq:dos_real}, and \eqref{eq:distribution_real}, we now collect the current terms by their powers of \(\tau\). The equilibrium contribution is
\begin{align}
    j^{\tau^0}_a
    =&-e\int\frac{\mathrm d^3p}{(2\pi\hbar)^3}
    \left[
    f_0(\varepsilon_n)v_a
    +\hbar^2G^{rr}_{ij}\left(
    \frac{\partial^2\varepsilon_n}{\partial p_i\partial p_j}v_a f_0(\varepsilon_n)
    +\frac{1}{2}v_i v_jv_a
    \frac{\partial f_0(\varepsilon_n)}{\partial\varepsilon_n}
    \right)\right]
    \nonumber\\
    =&-e\int\frac{\mathrm d^3p}{(2\pi\hbar)^3}
    \left[
    \frac{\partial F_0(\varepsilon_n)}{\partial p_a}
    +\hbar^2G^{rr}_{ij}\left\{
    \frac{\partial}{\partial p_i}\left(v_jv_a f_0\right)
    -\frac{1}{2}\frac{\partial}{\partial p_a}\left(v_iv_j f_0\right)\right\}
    \right],
    \label{eq:j_tau0_real}
\end{align}
where \(\mathrm dF_0/\mathrm d\varepsilon_n=f_0(\varepsilon_n)\). For a periodic Brillouin zone this total derivative does not contribute to the transport current, as expected for equilibrium.

The \(\tau^1\) contribution is
\begin{align}
    j^{\tau^1}_a
    =&-\tau e^2E_b\int\frac{\mathrm d^3p}{(2\pi\hbar)^3}
    \left[
    v_a v_b\frac{\partial f_0(\varepsilon_n)}{\partial\varepsilon_n}+\hbar^2G^{rr}_{ij}\left(\frac{\partial^2\varepsilon_n}{\partial p_i\partial p_j}
    \frac{\partial f_0(\varepsilon_n)}{\partial\varepsilon_n}+\frac{1}{2}v_i v_j\frac{\partial^2 f_0(\varepsilon_n)}{\partial\varepsilon_n^2}
    \right)v_a v_b
    \right] .
    \label{eq:j_tau1_real}
\end{align}
The first term is the ordinary Drude contribution. The second term is a metric correction to the longitudinal response; it is symmetric under \(a\leftrightarrow b\) and therefore does not generate a Hall response.

Finally, the \(\tau^2\) term is
\begin{equation}
    j^{\tau^2}_a
    =\tau^2e^2E_b\int\frac{\mathrm d^3p}{(2\pi\hbar)^3}
    \left(\hbar\Omega^{rr}_{cj}-\hbar^2\Gamma^{rrr}_{cjk}v_k\right)
    v_jv_a\frac{\partial^2\varepsilon_n}{\partial p_b\partial p_c}
    \frac{\partial f_0(\varepsilon_n)}{\partial\varepsilon_n} .
    \label{eq:j_tau2_real}
\end{equation}
This term can produce a Hall response. The \(\mathcal O(\hbar)\) term is the topological Hall effect arising from the real-space Berry curvature, while the \(\mathcal O(\hbar^2)\) term proportional to \(\Gamma^{rrr}\) is the emergent-gravity Hall contribution discussed in the main text.

\subsection{Momentum-space geometry}

We next consider the case in which the periodic part of the Bloch state depends on momentum but not on real-space texture. Using the same total-derivative convention for the canonical term, the Lagrangian is
\begin{equation}
    L=-\tilde\varepsilon-\dot{\bm p}\cdot\bm r
    +\hbar A^p_{nn,j}\dot p_j
    +\frac{\hbar^2}{2}G^{pp}_{ij}\dot p_i\dot p_j .
    \label{eq:Lagrangian_momentum}
\end{equation}
The equations of motion are
\begin{align}
    \begin{cases}
        \dot r_a=v_a-\hbar\Omega^{pp}_{ai}\dot p_i
        +\hbar^2\Gamma^{ppp}_{aij}\dot p_i\dot p_j
        +\hbar^2G^{pp}_{ai}\ddot p_i,\\
        \dot p_a=-eE_a.
        \label{eq:eom_momentum}
    \end{cases}
\end{align}
For a uniform electric field, \(\ddot p_i=0\). The phase-space correction is absent for a spatially uniform electric field~\cite{MamedaYamamoto}, so we set \(D=1\). To extract the intrinsic response through \(\mathcal O(\hbar^2E^2)\), it is sufficient to use the equilibrium distribution \(f_0(\varepsilon_n(\bm p))\). The resulting current is
\begin{align}
    j_a
    &=-e\int\frac{\mathrm d^3p}{(2\pi\hbar)^3}
    f_0(\varepsilon_n)
    \left[v_a+\hbar e\Omega^{pp}_{ai}E_i
    +\hbar^2e^2\Gamma^{ppp}_{aij}E_iE_j\right] .
    \label{eq:j_momentum_intrinsic}
\end{align}
The equilibrium part proportional to \(v_a\) vanishes after integration over a periodic Brillouin zone. The term linear in \(\bm E\) is the intrinsic anomalous Hall current. The first contribution from the Christoffel symbol appears at second order in the electric field, which includes the momentum-space counterpart of the emergent-gravity Hall response.

\section{Real-space gravity for a skyrmion}
We explicitly present the Christoffel symbols for a magnetic skyrmion. For the spin structure defined in Eqs.~(18) and (19) of the main text, the Christoffel symbols are given by
\begin{align}
    \Gamma^{rrr}_{x,xx}&=-\Gamma^{rrr}_{x,yy}=\Gamma^{rrr}_{y,xy}=\frac{-2 a^2 x}{h\left(a^2+x^2+y^2\right)^3},\label{eq:gam_xxx}\\
    \Gamma^{rrr}_{x,xy}&=-\Gamma^{rrr}_{y,xx}=\Gamma^{rrr}_{y,yy}=\frac{-2 a^2 y}{h\left(a^2+x^2+y^2\right)^3}\label{eq:gam_yyy}.
\end{align}
In momentum space, we consider a noncentrosymmetric magnet and assume 
$$h_0=\frac{\hbar^2\left|\bm{q}\right|^2}{2m}+\alpha q_x^3=\varepsilon_0\left(\tilde{q}_x^2+\tilde{q}_y^2+\tilde{\alpha}\tilde{q}_x^3\right),$$
where $m$ is the electron mass. For this model, the emergent-gravity Hall conductivity at zero temperature is given by
\begin{align}
    \frac{\sigma^{\mathrm{asym}}_{xy}}{e^2/\hbar}&=-\frac{\tilde{\tau}^2}{2\tilde{a}^3}\tilde{\Gamma}^{rrr}_{y,yy}\sum_{n\in\mathrm{occ}}\int\frac{\rmd^2\tilde{q}}{(2\pi)^2} \left\{\left(1+3\tilde{\alpha}\tilde{q}_x\right)^2\tilde{q}_x^2+\left(1+12\tilde{\alpha}\tilde{q}_x\right)\tilde{q}_y^2\right\}\left(1+3\tilde{\alpha}\tilde{q}_x\right)\tilde{q}_x\delta\left(\tilde{\mu}-\tilde{\varepsilon}_n\right),
\end{align}
where we have introduced the dimensionless quantities $\tilde{\Gamma}^{rrr}_{a,bc}=\varepsilon_0a^3\Gamma^{rrr}_{a,bc},\ \tilde{\tau}=\tau\varepsilon_0/\hbar,\ \tilde{a}=aq_0,\ \tilde{q}_i=q_i/q_0,\ q_0=\sqrt{2m\varepsilon_0}/\hbar,\ \tilde{\alpha}=\alpha q_0^3/\varepsilon_0,\ \tilde{\mu}=\mu/\varepsilon_0$, and $\tilde{\varepsilon}_n=\varepsilon_n/\varepsilon_0$, where $\varepsilon_0$ is a characteristic energy scale. In deriving this expression, we used the relations among the Christoffel symbols, Eqs.~\eqref{eq:gam_xxx} and \eqref{eq:gam_yyy}, and dropped terms containing odd powers of $q_y$, since they vanish upon integration. When we set $\tilde{\mu}=\mu/\varepsilon_0=0$, only the lower band with $\tilde{\varepsilon}_1=h_0/\varepsilon_0-h/\varepsilon_0$ contributes to the calculation. Although the energy $h_0$ is unbounded from below as $\tilde{q}_x\to-\infty$, we focus on the Fermi pocket around the origin, since this model is intended as a toy model around the origin in momentum space, and truncate the momentum region to $\tilde{q}_x,\tilde{q}_y\in[-2,2]$ for the calculation of Fig.~2 in the main text.

\section{Momentum-space gravity for two-dimensional system with Rashba-type spin-orbit interaction under a magnetic field}
We explicitly show Christoffel symbols for the Hamiltonian~(24) of the main text. This Hamiltonian has the form $H(\vb{q})=h_0\sigma_0+\vb{h}\cdot\boldsymbol{\sigma}$ with $h_0=\hbar^2\abs{\vb{q}}^2/2m$, $h_x=-Cq_y+H_x$, and $h_y=Cq_x+H_y$. From Eq.~(14) of the main text, the Christoffel symbols read
\begin{align}
    \Gamma^{qqq}_{x,xx}&=-\frac{5C^3h_x^2h_y}{8h^7}\label{eq:mom_Gam_xxx},\\
    \Gamma^{qqq}_{x,yy}&=\frac{C^3}{8h^4}\left(\frac{h_y}{h}+5\frac{h_yh_x^2}{h^3}\right),\\
    \Gamma^{qqq}_{y,xx}&=-\frac{C^3}{8h^4}\left(\frac{h_x}{h}+5\frac{h_xh_y^2}{h^3}\right),\\
    \Gamma^{qqq}_{y,yy}&=\frac{5C^3h_xh_y^2}{8h^7}\label{eq:mom_Gam_yyy}.
\end{align}
These quantities are divergent at $h\coloneqq\abs{\vb{h}(\vb{q})}=0$, where two bands cross. The Christoffel symbols in Eqs.~\eqref{eq:mom_Gam_xxx} and \eqref{eq:mom_Gam_yyy} induce the nonreciprocal longitudinal transport which has been investigated in Ref.~\cite{salaQuantumMetricElectrons2025a}.

\section{Expectation value}
\label{sec:expectation_value}

Here we calculate the expectation value of the derivative of the Hamiltonian
with respect to a parameter. The purpose of this section is to show explicitly
how the term containing the Christoffel symbol is generated at order
$\hbar^2$.

We consider a non-degenerate instantaneous band $n$ and write
\begin{equation}
    H(\lambda)\lvert u_m\rangle=\varepsilon_m(\lambda)\lvert u_m\rangle .
\end{equation}
Throughout this section repeated parameter indices are summed over, while band
indices are summed only when the summation symbol is explicitly written. We
use
\begin{equation}
    \partial_i\equiv\frac{\partial}{\partial\lambda_i},\qquad
    \varepsilon_{mn}\equiv \varepsilon_m-\varepsilon_n,\qquad
    A^i_{mn}\equiv \left\langle u_m\middle|i\partial_i u_n\right\rangle ,
\end{equation}
and, when no parameter index is shown,
\begin{equation}
    A_{mn}\equiv A^i_{mn}\dot{\lambda}_i .
\end{equation}
Also, we use the parallel-transport gauge along the trajectory,
\begin{equation}
    A_{mm}=A^i_{mm}\dot{\lambda}_i=0 ,
\end{equation}
for each instantaneous state. The final expression is gauge invariant.

The adiabatically corrected state used in Eq.~\eqref{eq:Bloch_hbar2} may be written as
\begin{align}
    \lvert u\rangle
    =&\left[
        1-\frac{\hbar^2}{2}\sum_{m\neq n}
        \frac{A^i_{nm}A^j_{mn}}{\varepsilon_{mn}^2}
        \dot{\lambda}_i\dot{\lambda}_j
    \right]\lvert u_n\rangle
    \nonumber\\
    &+\sum_{m\neq n}
    \left[
        \hbar\frac{A_{mn}}{\varepsilon_{mn}}
        +\hbar^2\sum_{\ell\neq n}
        \frac{A_{m\ell}A_{\ell n}}
        {\varepsilon_{mn}\varepsilon_{\ell n}}
        +\hbar^2\frac{i}{\varepsilon_{mn}}
        \frac{d}{dt}\left(\frac{A_{mn}}{\varepsilon_{mn}}\right)
    \right]\lvert u_m\rangle
    +O(\hbar^3).
    \label{eq:adiabatic_state_for_expectation}
\end{align}
The Feynman--Hellmann identity gives
\begin{equation}
    \left\langle u_\ell\middle|\partial_a H\middle|u_m\right\rangle
    =
    \delta_{\ell m}\,\partial_a\varepsilon_m
    -i\varepsilon_{m\ell}A^a_{\ell m}.
    \label{eq:FH_identity_expectation}
\end{equation}

We now expand
\begin{equation}
    \left\langle u\middle|\partial_a H\middle|u\right\rangle
    =
    \partial_a\varepsilon_n
    +\hbar X^{(1)}_a
    +\hbar^2 X^{(2)}_a
    +O(\hbar^3).
\end{equation}
The zeroth-order term is simply $\partial_a\varepsilon_n$. The first-order term comes from the interference between $\lvert u_n\rangle$ and the first-order correction to the state:
\begin{align}
    X^{(1)}_a
    &=
    \sum_{m\neq n}
    \left[
        \frac{A_{mn}}{\varepsilon_{mn}}
        \left(-i\varepsilon_{mn}A^a_{nm}\right)
        +\mathrm{c.c.}
    \right]
    \nonumber\\
    &=
    2\Im\left[
        \sum_{m\neq n}A^a_{nm}A^i_{mn}
    \right]\dot{\lambda}_i
    =
    -\Omega_{ai}\dot{\lambda}_i ,
    \label{eq:first_order_expectation}
\end{align}
where
\begin{equation}
    \Omega_{ai}
    =
    \partial_a A^i_{nn}-\partial_i A^a_{nn}
    =
    -2\Im\left[
        \sum_{m\neq n}A^a_{nm}A^i_{mn}
    \right]
\end{equation}
is the Berry curvature.

The order-$\hbar^2$ contribution contains four parts: the time derivative of the first-order coefficient, the product of two first-order coefficients, the normalization correction, and the explicit second-order coefficient in Eq.~\eqref{eq:adiabatic_state_for_expectation}.  We write them as
\begin{equation}
    X^{(2)}_a
    =
    T^{(\dot{A})}_a
    +T^{(1\times 1)}_a
    +T^{(N)}_a
    +T^{(2)}_a .
\end{equation}

First, the time derivative term gives
\begin{align}
    T^{(\dot{A})}_a
    &=
    \sum_{m\neq n}
    \left[
        \frac{i}{\varepsilon_{mn}}
        \frac{d}{dt}\left(\frac{A_{mn}}{\varepsilon_{mn}}\right)
        \left(-i\varepsilon_{mn}A^a_{nm}\right)
        +\mathrm{c.c.}
    \right]
    \nonumber\\
    &=
    2\Re\left[
        \sum_{m\neq n}
        A^a_{nm}
        \frac{d}{dt}
        \left(
            \frac{A^i_{mn}\dot{\lambda}_i}{\varepsilon_{mn}}
        \right)
    \right]
    \nonumber\\
    &=
    G_{ai}\ddot{\lambda}_i
    +
    2\Re\left[
        \sum_{m\neq n}
        A^a_{nm}
        \partial_j
        \left(
            \frac{A^i_{mn}}{\varepsilon_{mn}}
        \right)
    \right]\dot{\lambda}_i\dot{\lambda}_j ,
    \label{eq:TdotA}
\end{align}
where we have introduced the weighted quantum metric
\begin{equation}
    G_{ij}
    =
    2\Re\left[
        \sum_{m\neq n}
        \frac{A^i_{nm}A^j_{mn}}{\varepsilon_{mn}}
    \right].
    \label{eq:weighted_metric_expectation}
\end{equation}
The product of two first-order amplitudes is given by
\begin{align}
    T^{(1\times 1)}_a
    &=
    \sum_{\ell,m\neq n}
    \frac{A_{n\ell}A_{mn}}
    {\varepsilon_{\ell n}\varepsilon_{mn}}
    \left\langle u_\ell\middle|\partial_a H\middle|u_m\right\rangle
    \nonumber\\
    &=
    \left[
        -i
        \sum_{\substack{\ell,m\neq n\\ \ell\neq m}}
        \frac{\varepsilon_{m\ell}}
        {\varepsilon_{\ell n}\varepsilon_{mn}}
        A^i_{n\ell}A^j_{mn}A^a_{\ell m}
        +
        \sum_{m\neq n}
        \frac{A^i_{nm}A^j_{mn}}{\varepsilon_{mn}^2}
        \partial_a\varepsilon_m
    \right]\dot{\lambda}_i\dot{\lambda}_j .
    \label{eq:T11}
\end{align}
The normalization correction contributes to
\begin{equation}
    T^{(N)}_a
    =
    -
    \sum_{m\neq n}
    \frac{A^i_{nm}A^j_{mn}}{\varepsilon_{mn}^2}
    \partial_a\varepsilon_n\,
    \dot{\lambda}_i\dot{\lambda}_j .
    \label{eq:TN}
\end{equation}
Finally, the explicit second-order amplitude gives
\begin{align}
    T^{(2)}_a
    &=
    \sum_{m\neq n}
    \left[
        \sum_{\ell\neq n}
        \frac{A_{m\ell}A_{\ell n}}
        {\varepsilon_{mn}\varepsilon_{\ell n}}
        \left(-i\varepsilon_{mn}A^a_{nm}\right)
        +\mathrm{c.c.}
    \right]
    \nonumber\\
    &=
    2\Re\left[
        -i
        \sum_{\substack{m\neq n\\ \ell\neq n,m}}
        \frac{
            A^a_{nm}A^i_{m\ell}A^j_{\ell n}
        }{\varepsilon_{\ell n}}
    \right]\dot{\lambda}_i\dot{\lambda}_j .
    \label{eq:T2}
\end{align}
In the last line, the $\ell=m$ term is proportional to $A_{mm}=A^i_{mm}\dot{\lambda}_i$ and therefore vanishes in the parallel-transport gauge. Using
\begin{equation}
    \varepsilon_{m\ell}
    =
    \varepsilon_{mn}-\varepsilon_{\ell n},
\end{equation}
and relabelling dummy band indices, Eqs.~\eqref{eq:T11}--\eqref{eq:T2} combine into
\begin{align}
    T^{(1\times 1)}_a+T^{(N)}_a+T^{(2)}_a
    &=
    2\Re\left[
        i
        \sum_{\substack{m\neq n\\ \ell\neq n,m}}
        \frac{
            A^i_{n\ell}
            \left(
                A^j_{\ell m}A^a_{mn}
                -
                A^a_{\ell m}A^j_{mn}
            \right)
        }{\varepsilon_{\ell n}}
    \right]\dot{\lambda}_i\dot{\lambda}_j
    \nonumber\\
    &\quad
    +
    \sum_{m\neq n}
    \frac{A^i_{nm}A^j_{mn}}{\varepsilon_{mn}^2}
    \partial_a\varepsilon_{mn}\,
    \dot{\lambda}_i\dot{\lambda}_j .
    \label{eq:triple_sum_before_identity}
\end{align}
The double sum is simplified by differentiating the Berry connection:
\begin{equation}
    \partial_j A^a_{\ell n}
    -
    \partial_a A^j_{\ell n}
    =
    i\sum_r
    \left(
        A^j_{\ell r}A^a_{rn}
        -
        A^a_{\ell r}A^j_{rn}
    \right).
    \label{eq:connection_derivative_identity}
\end{equation}
The diagonal terms of $A$ in the sum over $r$ are harmless: one is antisymmetric under $i\leftrightarrow j$ after taking the real part and therefore vanishes upon contraction with $\dot{\lambda}_i\dot{\lambda}_j$, while the other is proportional to $A^j_{nn}\dot{\lambda}_j=0$ in the parallel-transport gauge. Hence Eq.~\eqref{eq:triple_sum_before_identity} becomes
\begin{align}
    T^{(1\times 1)}_a+T^{(N)}_a+T^{(2)}_a
    &=
    2\Re\left[
        \sum_{m\neq n}
        \frac{
            A^i_{nm}
            \left(
                \partial_j A^a_{mn}
                -
                \partial_a A^j_{mn}
            \right)
        }{\varepsilon_{mn}}
    \right]\dot{\lambda}_i\dot{\lambda}_j
    \nonumber\\
    &\quad
    +
    \sum_{m\neq n}
    \frac{A^i_{nm}A^j_{mn}}{\varepsilon_{mn}^2}
    \partial_a\varepsilon_{mn}\,
    \dot{\lambda}_i\dot{\lambda}_j .
    \label{eq:T11_TN_T2_simplified}
\end{align}
Combining Eq.~\eqref{eq:TdotA} with Eq.~\eqref{eq:T11_TN_T2_simplified}, the coefficient of $\dot{\lambda}_i\dot{\lambda}_j$ can be written as
\begin{align}
    &2\Re\left[
        \sum_{m\neq n}
        A^a_{nm}
        \partial_j
        \left(
            \frac{A^i_{mn}}{\varepsilon_{mn}}
        \right)
    \right]
    +
    2\Re\left[
        \sum_{m\neq n}
        \frac{
            A^i_{nm}
            \left(
                \partial_j A^a_{mn}
                -
                \partial_a A^j_{mn}
            \right)
        }{\varepsilon_{mn}}
    \right]
    +
    \sum_{m\neq n}
    \frac{A^i_{nm}A^j_{mn}}{\varepsilon_{mn}^2}
    \partial_a\varepsilon_{mn}
    \nonumber\\
    &\hspace{2cm}
    =
    \partial_j G_{ai}
    -
    \frac{1}{2}\partial_a G_{ij}.
    \label{eq:Christoffel_intermediate}
\end{align}
Since this expression is contracted with the symmetric tensor
$\dot{\lambda}_i\dot{\lambda}_j$, we may symmetrize the first term in
$i$ and $j$:
\begin{equation}
    \left(
        \partial_j G_{ai}
        -
        \frac{1}{2}\partial_a G_{ij}
    \right)
    \dot{\lambda}_i\dot{\lambda}_j
    =
    \Gamma_{a,ij}\dot{\lambda}_i\dot{\lambda}_j ,
\end{equation}
where
\begin{equation}
    \Gamma_{a,ij}
    =
    \frac{1}{2}
    \left(         \partial_i G_{aj}
        +
        \partial_j G_{ai}
        -         \partial_a G_{ij}
    \right)
\end{equation}
is the Christoffel symbol of the first kind associated with the weighted metric
$G_{ij}$.

Therefore,
\begin{equation}
    X^{(2)}_a
    =
    G_{ai}\ddot{\lambda}_i
    +
    \Gamma_{a,ij}\dot{\lambda}_i\dot{\lambda}_j .
\end{equation}
Collecting all the orders, we obtain
\begin{equation}
    \left\langle u\middle|
    \frac{\partial H}{\partial\lambda_a}
    \middle|u\right\rangle
    =
    \frac{\partial\varepsilon_n}{\partial\lambda_a}
    -
    \hbar\Omega_{ai}\dot{\lambda}_i
    +
    \hbar^2
    \left(
        \Gamma_{a,ij}\dot{\lambda}_i\dot{\lambda}_j
        +
        G_{ai}\ddot{\lambda}_i
    \right)
    +O(\hbar^3).
    \label{eq:expectation_value_final}
\end{equation}
Choosing $\lambda_i=p_i$ reproduces the momentum-space equation of motion, Eq.~\eqref{eq:eom_momentum}.

\section{Charge pumping by parameter space gravity}
Here, we consider the dependence on an external parameter $\lambda$ and show that Christoffel symbols also appear in charge pumping in insulators.

In insulating systems, only the $\tau^0$ terms in the Boltzmann equation contribute to the current. The terms involving $\dot{\lambda}$ are  
\begin{align}
    j_{a}\propto\hbar\int f_0\left\{\qty(-\frac{1}{\hbar}\Omega^{q\lambda}_{a}+2\Gamma^{qr\lambda}_{a,i\phantom{j}}\dot{r}_i+2\Gamma^{qq\lambda}_{a,i\phantom{j}}\dot{q}_i+2\Gamma^{q\lambda t}_{a,})\dot{\lambda}+\Gamma^{q\lambda\lambda}_{a,}\dot{\lambda}\dot{\lambda}+2G^{q\lambda}_{a}\ddot{\lambda}\right\}.
\end{align}
In Ref.~\cite{xiao_polarization_2009}, adiabatic charge pumping driven by the Berry curvatures was investigated. Here, by incorporating nonadiabatic processes, we find that charge currents can also arise from Christoffel symbols in parameter space.

% \bibliography{EG.bib}

% \end{document}

\end{document}